\documentclass[aps,prd,preprint,superscriptaddress,tightenlines,nofootinbib]{revtex4}
\usepackage{graphicx}
\usepackage{dcolumn}
\usepackage{bm}
\begin{document}
\preprint{CLNS 07/1998}       
\preprint{CLEO 07-07}         

\title{Measurement of the total hadronic cross section in $e^{+}e^{-}$ annihilation below 10.56 GeV}
\author{D.~Besson}
\affiliation{University of Kansas, Lawrence, Kansas 66045}
\author{T.~K.~Pedlar}
\affiliation{Luther College, Decorah, Iowa 52101}
\author{D.~Cronin-Hennessy}
\author{K.~Y.~Gao}
\author{J.~Hietala}
\author{T.~Klein}
\author{Y.~Kubota}
\author{B.~W.~Lang}
\author{R.~Poling}
\author{A.~W.~Scott}
\author{A.~Smith}
\author{P.~Zweber}
\affiliation{University of Minnesota, Minneapolis, Minnesota 55455}
% ...
\collaboration{CLEO Collaboration}
\noaffiliation

\begin{abstract}
{ Using the CLEO~III  detector, we measure absolute cross sections for  
$e^+e^-\rightarrow \rm{hadrons}$ at seven center-of-mass energies 
between 6.964 and 10.538~GeV. The values of $R$, the ratio of hadronic 
and muon pair production cross sections, are determined within 2\% 
total r.m.s. uncertainty.}
\end{abstract}
\maketitle
\section{Introduction}
\label{introduction}
The total cross section for hadron production in electron-positron annihilation
is often expressed in terms of the ratio $R(s)$ of the radiation-corrected
measured hadronic cross section to the calculated lowest-order
cross section for muon pair production.  That is,
\begin{eqnarray} 
R(s) = \sigma_{0}(e^+e^-\rightarrow \rm{hadrons} )/\sigma_{0}(e^+e^-\rightarrow \mu^+\mu^-),
\label{defR}
\end{eqnarray}
where $\sigma_0(e^+e^-\to\mu^+\mu^-)=4\pi\alpha^2/3s$ . 
In the simplest form of the quark model $R(s)$ is $R_0(s)$, the sum of the 
squares of the charges for the quark species allowed by energy 
conservation, including a factor of $N_{c}=3$ for the colors.   For energies 
between $c\bar{c}$ and $b\bar{b}$ thresholds, 
$R_0=3\Sigma_{udsc}Q_{f}^2=10/3$, where $Q_{f}$ represents the charge of 
the quark $f=u,d,s,c$.
 
The strong interactions of the quarks can be accounted for in a perturbation
theory expansion in the QCD coupling $\alpha_s(s)$.  In the modified 
minimal-subtraction scheme at the four-loop level \cite{Surguladze}:
\begin{eqnarray}
R(s) & = & R_0\left[ 1 + C_1\frac{\alpha_{s}(s)}{\pi} + C_2\left(\frac{\alpha_{s}(s)}{\pi}\right)^2  \right. \nonumber \\
  &   & \left. \mbox{}+ C_3\left(\frac{\alpha_{s}(s)}{\pi}\right)^3 + O(\alpha_s^4(s))\right]. 
\label{FormR}
\end{eqnarray}
For the four-flavor case the coefficients are  $C_1=1$, $C_2=1.525$, and 
$C_3=-11.686$ \cite{Surguladze},~\cite{Davier}. 
Asymptotic freedom in QCD implies that the coupling $\alpha_s$ is not a
constant but a smoothly decreasing function of $s$.  The aim of the present
experiment is to use measurements of $R(s)$ to make an accurate test of
this relation in the 48.5 to 111.0~$\rm{GeV}^2$ region of $s$,
where $u,d,s,$ and $c$ quarks are produced.

We use data collected at seven center-of-mass energies at the Cornell
Electron Storage Ring (CESR).  Electrons and positrons are stored at equal 
energies in oppositely rotating 
orbits in the Cornell Electron Storage Ring (CESR), and they collide at a
2~mrad angle at the center 
of the CLEO~III detector \cite{CLEO-Detector}.    
Charged tracks are reconstructed
in a 4-layer silicon strip detector and a 47-layer wire drift chamber, 
and their momenta are determined from their radii
of curvature in a 1.5~T magnetic field created by a superconducting solenoid
with its axis aligned along the average of the two colliding beam lines.  
The CsI scintillator shower calorimeter forms a cylindrical barrel around the 
tracking volume, covering polar angles $\theta$ with respect to the beam axis 
with $\mid \cos \theta \mid<0.85$, supplemented by endcaps extending 
the range to 
$\mid \cos \theta \mid<0.93$.  Electromagnetic showers are detected with an r.m.s. 
resolution of about 75~MeV at 5~GeV.

\section{ANALYSIS OVERVIEW}
\label{ANALYSIS-OVERVIEW}
To obtain the continuum hadronic annihilation cross section referred to in 
Eqs.~(\ref{defR},~\ref{FormR}), we must 
correct the observed yield for detection efficiency, for 
initial state radiation including its influence on efficiency, for other 
higher order QED effects, and for contributions from the radiative 
tails of lower energy resonances.  We write the observed continuum hadronic 
cross section as the sum of $\sigma_{\mbox{\scriptsize sv}}$ from soft photon radiation
and virtual higher order diagrams, 
plus $\sigma_{\mbox{\scriptsize hard}}$ from hard photon radiation, plus
$\sigma_{\mbox{\scriptsize res}}$ from radiative tails of lower mass
$q\bar{q}$ bound states 
\cite{BehrendsKleiss, BonneauMartin, VP, SLAC-4160}:
\begin{eqnarray}
\sigma_{\mbox{\scriptsize obs}}(s) = \sigma_{\mbox{\scriptsize sv}}(s) +  \sigma_{\mbox{\scriptsize hard}}(s) + \sigma_{\mbox{\scriptsize res}}(s).
\label{sigmaObs}
\end{eqnarray}
The sum of the soft and virtual term can be expressed as
\begin{eqnarray}
\sigma_{\rm sv}(s) = \epsilon(0)\sigma_{0}(s)\delta_{\rm sv},
\label{FormSV}
\end{eqnarray}
where $\epsilon(0)$ is the detection efficiency for events with no  
initial state radiation, 
$\sigma_0$ is the desired corrected hadronic Born cross
section, and $\delta_{\mbox{\scriptsize sv}}=\delta_0+\delta_{\mbox{\scriptsize vp}}$ accounts for 
soft photon emission and hadronic and leptonic vacuum polarization: 
\begin{eqnarray}
\delta_0=\frac{2\alpha}{\pi}\left(\frac{3}{4}\ln\frac{s}{m_e^2}+\frac{\pi^2}{6} -1\right),
\end{eqnarray}
and
\begin{eqnarray} 
\delta_{\rm vp}=\sum_{f}\frac{2\alpha}{\pi}\left(\frac{1}{3}\ln\frac{s}{m_f^2}-\frac{5}{9}\right),
\label{deltaVP}
\end{eqnarray}
where  $f=e,\mu, \tau, h $. A more accurate approximation of hadronic vacuum polarization can 
be found in reference \cite{Pietrzyk}.

The hard photon piece involves an integral of the continuum cross section over 
all radiated energies \cite{first20}:
\begin{eqnarray}
\sigma_{\mbox{\scriptsize hard}}(s)=\sigma_0(s)I_{\mbox{\scriptsize hard}},
\label{sigmaHard-sigma0}
\end{eqnarray}
with
\begin{eqnarray}
I_{\mbox{\scriptsize hard}} & = & \int_0^{k_{\mbox{\scriptsize max}}}\epsilon(k)\left( \frac{\sigma_{0}^{\mbox{\scriptsize cont}}(s^{\prime})}{\sigma_0^{\mbox{\scriptsize cont}}(s)}\right) \times \nonumber \\
        &   & \frac{t}{k^{1-t}} \left(1-k+\frac{k^2}{2}\right)dk,
\label{FormHard}
\end{eqnarray}
where $k=(s-s^{\prime})/s$  is the normalized radiated energy,
$\sqrt{s^{\prime}}$ is the center of mass energy after the initial state radiation, and
the number of equivalent radiation lengths is
\begin{eqnarray}
t=\frac{2\alpha}{\pi}\left(\ln\frac{s}{m_e^2}-1\right).
\end{eqnarray}

Before making corrections to the observed continuum hadronic cross section, 
we first have to subtract the
contributions from lower energy $q\bar{q}$ resonances excited through
the radiative process $e^+e^-\to\gamma q\bar{q}$ (the last term in
Eq.~(\ref{sigmaObs})).
For each contributing
resonance the corresponding term in $\sigma_{\mbox{\scriptsize res}}$ is given 
by the product of the detection efficiency $\epsilon(k)$, the non-radiative cross section
energy integral (given in terms of the resonance coupling $\Gamma_{ee}$
to $e^+e^-$, the resonance hadronic branching ratio $B_{\rm had}$, and the resonance mass M), 
and the radiative kernel, $F(k,s)$:
\begin{eqnarray}
\sigma_{\mbox{\scriptsize res}}(s) = \sum\epsilon(k)\left(\frac{6\pi^2}{M^2}\Gamma_{ee}B_{\mbox{\scriptsize had}}\right)F(k,s),
\label{FormRes}
\end{eqnarray}
where the sum runs over all known low energy resonances which contributes observed hadronic cross section.
In lowest order the kernel  can be written as
\begin{eqnarray}
F(k,s)	\approx \frac{2M}{s}\frac{t}{k^{1-t}}\left(1-k+\frac{k^2}{2}\right),   
\label{FormResKernel}
\end{eqnarray}
with $k=(s-s^\prime)/s=1-M^2/s$.
A higher order representation of $F(k,s)$
can be found in reference~\cite{Kuraev}.

The goal of the measurement, the 
Born cross section, is derived from Eqs.~(\ref{sigmaObs}), (\ref{FormSV}), 
and (\ref{sigmaHard-sigma0}),
\begin{eqnarray}
\sigma_0(s) = \frac{\sigma_{\mbox{\scriptsize obs}}(s)-\sigma_{\mbox{\scriptsize res}}(s)}{\epsilon(0)\delta_{\mbox{\scriptsize sv}}+I_{\mbox{\scriptsize hard}}}.
\label{FormCross}
\end{eqnarray}
It is obtained from the observed hadronic cross section, 
$\sigma_{\rm obs}=(N_{\rm had}-N_{\rm bkg})/(\epsilon_{\rm trigger} \times \int{\cal{L}}dt)$, where $N_{\rm had}$ is the
number of detected events passing hadronic event selection criteria,
$N_{\rm bkg}$ is the number of background events passing the same requirements, $\epsilon_{\rm trigger}$
is the trigger efficiency correction, and
$\int{\cal{L}}dt$ is the integrated luminosity.
The background includes  $e,\mu$, and $\tau$ pairs, 
virtual photon-photon (two-photon) interactions ($e^+e^-\to e^+e^-\gamma^*\gamma^*\to e^+e^- \rm{~hadrons}$), 
beam-gas (beam particle interactions with residual gas molecules)
and beam-wall interactions (beam particle interacts with the 
beam pipe material), and cosmic rays.

\section{DATA SAMPLES}
\label{DATA-SAMPLE}
The center-of-mass energies $\sqrt{s}$ and integrated luminosities
$\int{\cal{L}}dt$ used here are listed in Table~\ref{tab:lumi}.  
\begin{table}[htbp]
\caption{Center-of-mass energies and integrated luminosities.}
\label{tab:lumi}
\begin{center}
\begin{tabular}{cc}
\hline\hline
 Energy  & Luminosity  \\
  GeV         & $\rm{pb}^{-1}$ \\
\hline
  10.538     & $904.50 \pm 0.30 \pm 9.00 $ \\
  10.330     & $149.80 \pm 0.10 \pm 1.60 $ \\
   9.996     & $432.60 \pm 0.20 \pm 4.80 $ \\
   9.432     & $183.00 \pm 0.10 \pm 2.00 $ \\
   8.380     & $~~~6.78 \pm 0.02 \pm 0.06 $ \\
   7.380     & $~~~8.48 \pm 0.02 \pm 0.07 $ \\
   6.964     & $~~~2.52 \pm 0.01 \pm 0.02 $ \\
\hline\hline
\end{tabular}
\end{center}
\end{table}

The luminosities for these
data were calculated from
measured rates of three different
QED processes, 
$e^+e^- \rightarrow e^+e^-,~\gamma\gamma,{\rm~and~} \mu^+\mu^-$,
using predicted cross sections from
the Babayaga MC generator \cite{Babayaga},
with selection criteria similar to
those in Ref.~\cite{CLEOIII:Lumi}. Particular
care was taken to 
include
the effects of $\Upsilon(nS)\rightarrow l^+l^-$, where $l$ 
stands for lepton: $e$ or $\mu$, 
from either direct production
or with an intermediate radiative return 
\cite{Kuraev}.
The resulting luminosities from
all three reactions were found to be
consistent within their systematic
uncertainties of 1.8\%, 1.2\%, and 1.2\%
for the $\gamma\gamma$, $e^+e^-$, and  $\mu^+\mu^-$
channels, respectively, and a
single value at each energy was
obtained by combining the three values
according to their relative total uncertainties.
The total relative uncertainty on the combined luminosity
varied from 0.9-1.1\%, depending upon energy.

To study efficiencies and backgrounds we used GEANT-based 
Monte Carlo simulations \cite{GEANT321},
Jetset~7.3 \cite{Jetset} for hadronic events, and KORALB \cite{KORALB} for 
$\tau$ pairs.

\section{EVENT SELECTION AND DETECTION EFFICIENCIES}
\label{EVENT-SELECTION}
The detector trigger used in the measurement of the hadronic event
rate is the logical ``or'' of the following two sets of requirements: 
1) at least three axial-projection charged tracks 
(tracks which are detected in the inner 16 layers of the drift chamber) 
and at least one low-threshold electromagnetic shower; 
2) at least two stereo-projection charged tracks 
(tracks which are detected in the outer 31 layers of the drift chamber) 
and either at least two low-threshold  showers 
or at least one medium-threshold shower. 
The low (medium) trigger threshold for showers is approximately 
150~MeV (750~MeV).  We evaluate the efficiency by comparing rates (using
the hadronic selection criteria described in the next section) with
those obtained in special runs with a much looser detection trigger, namely
requiring two or more axial-projection charged tracks.  
Table~\ref{tab:effandtrig} shows the 
resulting trigger efficiency values, $\epsilon_{\rm trigger}$,  
for the various energy points.  The systematic uncertainties are 
conservatively taken as the deviations from 100\% efficiency.

\begin{table}[htbp]
\caption{Trigger efficiency correction $\epsilon_{\rm trigger}$, 
and net hadronic event detection efficiency $\epsilon(0)$ evaluated for 
the case of zero initial state radiation energy ($k=0$). 
The uncertainties are statistical only.}
\label{tab:effandtrig}
\begin{center}
\begin{tabular}{ccc}
\hline\hline 
Energy    & $\epsilon_{\rm trigger}$ & $\epsilon(0)$ \\
 GeV      &   \%          &     \%        \\
\hline
10.538     & $~~99.91 \pm 0.01$ & $~~87.43 \pm 0.01 $ \\
10.330     & $~~99.91 \pm 0.01$ & $~~87.42 \pm 0.03 $ \\
 9.996     & $~~99.89 \pm 0.01$ & $~~87.13 \pm 0.03 $ \\
 9.432     & $~~99.92 \pm 0.01$ & $~~86.05 \pm 0.04 $ \\
 8.380     & $~~99.88 \pm 0.02$ & $~~84.71 \pm 0.10 $ \\
 7.380     & $~~99.87 \pm 0.02$ & $~~82.95 \pm 0.10 $ \\
 6.964     & $~~99.81 \pm 0.03$ & $~~82.07 \pm 0.14 $ \\ 
\hline\hline
\end{tabular}
\end{center}
\end{table}

In order to suppress backgrounds from events other than $e^+e^- \to $~hadrons,
we apply selection requirements (cuts) to individual tracks and showers
as well as to entire events.  We now describe these cuts and
explain how we estimate their efficiency and associated uncertainties.
	
Table~\ref{tab:trackandshower} lists the requirements for accepting 
individual tracks and showers. 
\begin{table}[htbp]
\caption{Cuts used to select tracks and showers.}  
\label{tab:trackandshower}
\begin{center}
\begin{tabular}{ c l }
\hline\hline
 Variable  & Allowed range \\
\hline
$\chi^2$/NDF          &   $~~~<~100.0$ \\
hit fraction          &   $~~~(0.5,1.2)$ \\
$\mid d_0 \mid$       &   $~~~<~3.0$~cm \\
$\mid z_0 \mid $      &   $~~~<~18.0$~cm \\
error of $z_0$        &   $~~~<~25.0$~cm \\
$\mid \cot(\theta) \mid$    &   $~~~<~3.0424$ \\
error of $\cot(\theta)$ &  $~~~<~0.50$\\
 $P_{\rm track}/E_{\rm beam}$     &   $~~~(0.01,1.5)$  \\
 $E_{\rm shower}/E_{\rm beam}$      &   $~~~>0.01 $ \\
\hline\hline
\end{tabular}
\end{center}
\end{table}
We accept all tracks which have $\chi^2$ per degree of freedom 
$(\chi^2/{\rm NDF})$  less than 100.
Consistency with the beam collision point is enforced by the cut on $d_0$,
the distance of closest approach of the reconstructed track relative to
the beam axis, and on $z_0$, the distance between that point and the average
collision point on the beam axis.
The hit fraction is the ratio of the number of 
detected and expected hits when a track passes through the drift chamber,
and the $\theta$ is the polar angle of the track relative to the beam axis.  
We require that each track candidate 
carry at least 1\% of the beam momentum. To avoid 
poorly measured tracks, we require that the ratio of the track momentum 
to the beam momentum  be less than 1.5. We require that each accepted 
shower carry at least 1\% of the beam energy and that no track be 
associated with it.
  
The event selection criteria are listed in Table~\ref{tab:cuts}, and their 
description are provided in the following paragraphs. 
\begin{table}[htbp]
\caption{Hadronic event selection criteria.}  
\label{tab:cuts}
\begin{center}
\begin{tabular}{ c l }
\hline\hline
 Variable  & Allowed range \\
\hline
$\mid Z_{\mbox{\scriptsize vertex}} \mid $ &   $ ~~~< 6.0$~cm  \\
$E_{\mbox{\scriptsize vis}}/(2E_{\mbox{\scriptsize beam}})$     &   $ ~~~>0.5 $ \\
$\mid P_{\mbox{\scriptsize z}}^{\mbox{\scriptsize miss}}/E_{\mbox{\scriptsize vis}} \mid $       &   $ ~~~< 0.3$\\
$H_2/H_0   $               &   $~~~< 0.9 $\\
$E_{\mbox{\scriptsize cal}}/(2E_{\mbox{\scriptsize beam}})$     &   $ ~~~(0.15,0.9) $\\
$E_{\gamma}^{\mbox{\scriptsize max}}/E_{\mbox{\scriptsize beam}}$   &   $~~~<0.8$  \\
$N_{\mbox{\scriptsize ChargedTrack}}$       &   $~~~\ge 4$  \\
\hline\hline
\end{tabular}
\end{center}
\end{table}

The event $Z_{\rm vertex}$ is the weighted average of the $z_0$'s of the
charged tracks, and weights are defined as $1/\sigma^2_{z_0}$, where 
$\sigma_{z_0}$ is the uncertainty of the $z_0$.  
The purpose of this cut was to suppress background 
due to beam-gas, beam-wall collisions and 
from cosmic rays.  The measured $R$ value was 
sensitive at the level of only 0.16\% to the cut over the range from 
4 to 10~cm. 
\begin{figure}[htbp]
\includegraphics*[width=3.4in]{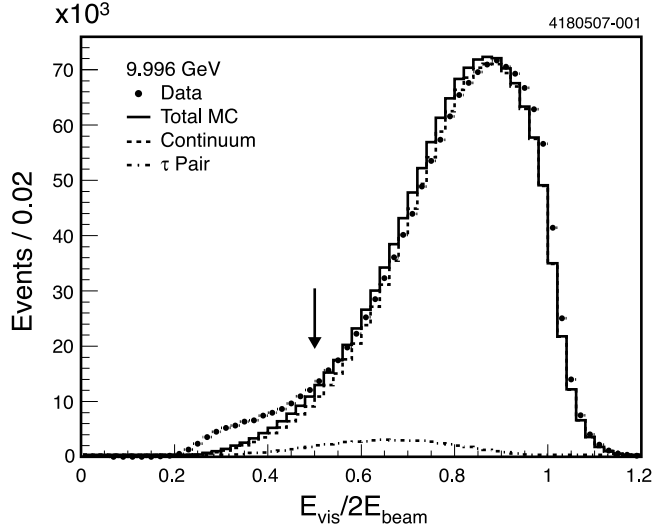}
\caption{Visible energy distribution at 9.996~GeV, normalized to 
center-of-mass energy, when all cuts are applied except
the one shown. In this as well as in all other figures, arrows
indicate cut values.}
\label{FigEvis_Ups2sOff}
\end{figure}

The visible energy $E_{\mbox{\scriptsize vis}}$ is defined as the sum of 
charged particle energies computed from the track momenta assuming pion masses,
plus the neutral shower energies measured in the electromagnetic
calorimeter.  We normalize the visible energy to the center-of-mass 
energy so that we can use the same cut for all runs.  
The visible energy cut is designed to suppress background 
from virtual photon-photon collisions and from beam-gas interactions.
Figure~\ref{FigEvis_Ups2sOff} shows the normalized visible energy  
distribution at 9.996~GeV.
We varied the cut in scaled visible energy ($E_{\rm vis}/(2E_{\rm beam})$) 
from its nominal 
level of 0.5 down to 0.3 and to 0.4 to estimate our sensitivity to the 
two-photon background. We also varied the 
cut upwards to 0.6 and to 0.7 to estimate the
discrepancy between data and Monte Carlo.

\begin{figure}[htbp]
\includegraphics*[width=3.4in]{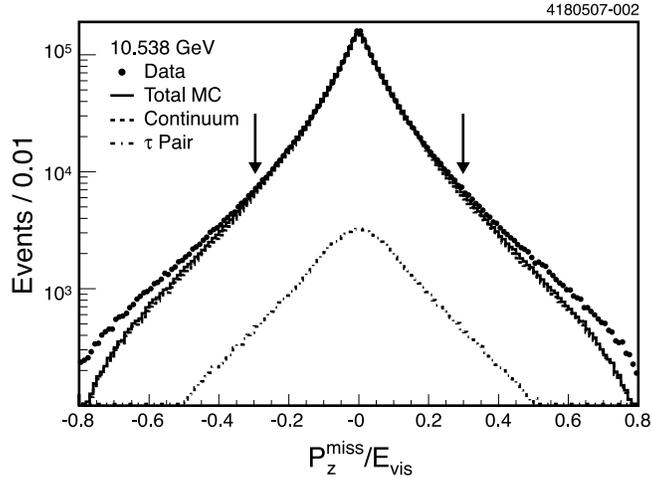}
\caption{The ratio of the z-component of the event missing momentum to 
event visible energy 
$P_{z}^{\mbox{\scriptsize miss}}/E_{\mbox{\scriptsize vis}}$ 
at 10.538~GeV.
}
\label{FigPzMiss_Ups4sOff}
\end{figure}

The background from 
two-photon and beam-gas events was also 
suppressed by  the cut on the ratio 
$\mid P_{z}^{\mbox{\scriptsize miss}}/E_{\mbox{\scriptsize vis}}\mid $ 
of the $z$-component of the event missing 
momentum and visible energy (see Fig.~\ref{FigPzMiss_Ups4sOff}).  The $R$ results 
varied by less than 0.65\% over the cut range 0.2 to 0.4.

\begin{figure}[htbp]
\includegraphics*[width=3.4in]{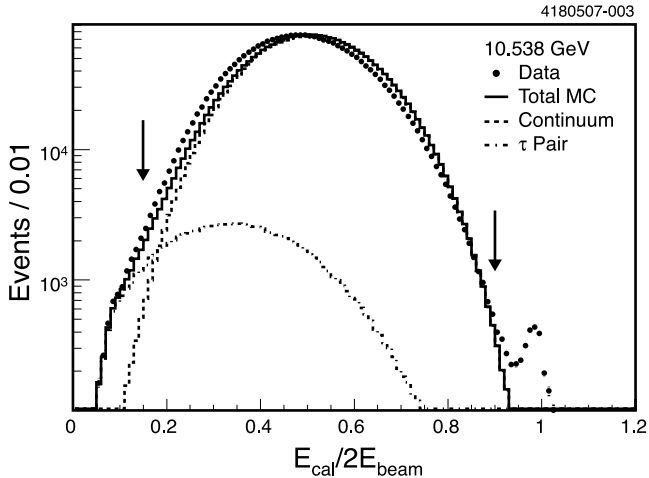}
\caption{The ratio of calorimeter energy of charged and neutral particles to 
$2E_{\rm beam}$ at 10.538~GeV.
}
\label{FigCE_Ups4sOff}
\end{figure}

The typical signature of a QED event ($e, \mu,$ or $\tau$ pair) is low charged track
multiplicity which would fail the four-track requirement, although higher 
multiplicity 
decays of taus are likely to pass. We rely on
calorimeter information to suppress the remaining background further.
The distribution of the ratio of calorimeter energy $E_{\mbox{\scriptsize cal}}$, 
which includes isolated showers as well as track matched showers, 
to the center-of-mass energy, is plotted in Fig.~\ref{FigCE_Ups4sOff}. 
It shows a contribution at low values from tau-pairs
and, at $E_{\mbox{\scriptsize cal}}/(2E_{\rm beam})=1$, 
a contribution from Bhabha-scattered electron pairs.  
The cuts eliminated both extremes.  Varying the upper cut from 0.8 to $\infty$
made less than 0.49\% difference in $R$ values.

\begin{figure}[htbp]
\includegraphics*[width=3.4in]{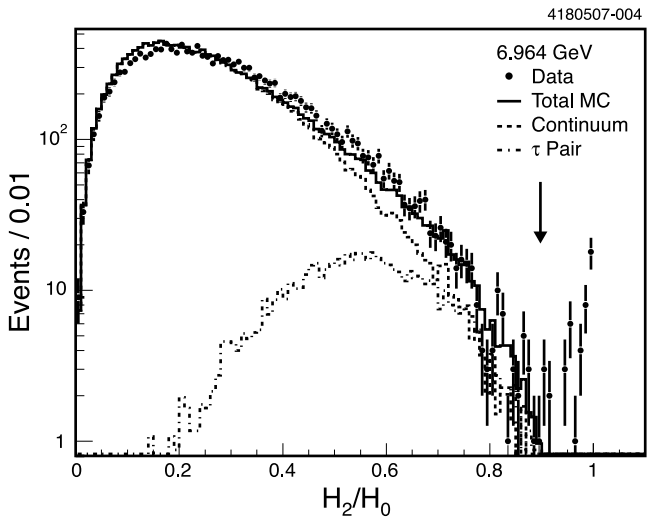}
\caption{Distribution of the ratio of Fox-Wolfram moments $H_2/H_0$ at 6.964~GeV. 
}
\label{FigR2_E70}
\end{figure}

Figure~\ref{FigR2_E70} shows the distribution of the ratio of Fox-Wolfram moments 
$H_2/H_0$ \cite{FoxWolfram}. The QED process $e^+e^- \rightarrow l^+l^-$, where
$l=e,\mu{\rm ~or~}\tau$, results
in a high value of this ratio. Cut variation above 0.8 produced 
a variation in $R$ of less than 0.54\%.
  
\begin{figure}[htbp]
\includegraphics*[width=3.4in]{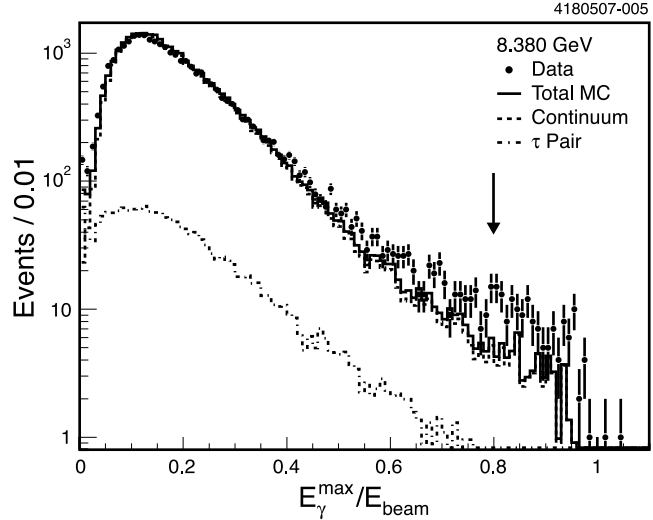}
\caption{The ratio of the most energetic shower energy to the beam energy at 8.380~GeV. 
}
\label{FigXg_E84}
\end{figure}

We also studied the ratio of the energy of the most energetic unmatched shower to the beam energy, 
shown in Fig.~\ref{FigXg_E84}. To suppress production 
of hadronic events at low center-of-mass energies through initial state radiation, we cut at 0.80. 
Varying the cut from 0.6 to infinity changed the value of $R$ less than 0.40\%.

\begin{figure}[htbp]
\includegraphics*[width=3.4in]{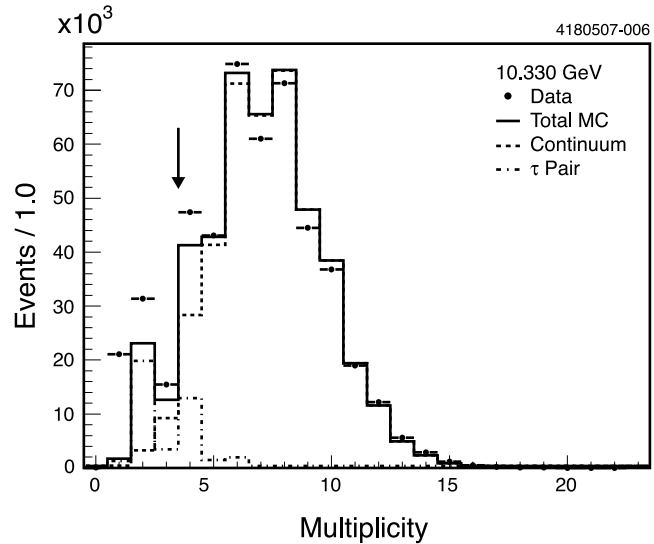}
\caption{Charged particle multiplicity distribution at 10.330~GeV.
}
\label{FigMult_Ups3sOff}
\end{figure}

The charged track multiplicity cut also suppressed QED events in our 
hadronic sample.  The multiplicity distribution 
(see Fig.~\ref{FigMult_Ups3sOff}) showed a significant
sensitivity of the accepted rate to the chosen cut level. However, there is a 
discrepancy between data and the Monte Carlo simulation. The discrepancy at
lower multiplicity values can be explained by the contribution of QED events;
the discrepancy at higher multiplicity values can be explained by inaccuracies at the Monte
Carlo generator level.

The selection criteria defined in Tables~\ref{tab:trackandshower}
and~\ref{tab:cuts} were used to determine the number of hadronic 
events and the hadronic detection efficiencies
for the various incident beam energies, 
using our  data and Monte Carlo samples (see Tables~\ref{tab:effandtrig} and~\ref{tab:res}).
We performed the efficiency calculation using  
weights assigned to each continuum multiplicity value instead of relying on the 
continuum Monte Carlo prediction for the multiplicity distribution. 
The weights were assigned by comparing data multiplicity 
with continuum and $\tau$-pair Monte Carlo samples:
\begin{eqnarray}
w_{n} = \frac{N_{n}^{\rm data} - 
N_{n}^{\tau-{\rm pair}}}{N_n^{\rm cont}},
\label{form-weight}
\end{eqnarray}
where $w_{n}$ is the weight which is assigned to a continuum event with 
multiplicity $n$; $N_{n}^{\rm data}$, $N_{n}^{\tau-\rm pair}$, and 
$N_n^{\rm cont}$ are the number of events in  the data sample,  $\tau$-pair, 
and continuum Monte Carlo samples with multiplicity $n$, respectively. Since
low multiplicity values are contaminated by $e^+e^-$ and $\mu^+\mu^-$ events, 
we assign a weight of 1.0 for multiplicities less than three.  

\begin{figure}[htbp]
\includegraphics*[width=3.4in]{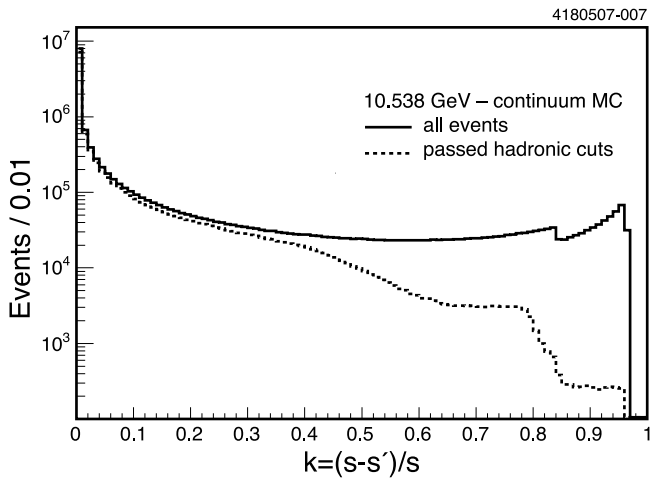}
\caption{$k$ distribution of the Monte Carlo continuum sample at 10.538 GeV. }
\label{FigISR_Ups4sOff}
\end{figure}

\begin{figure}
\includegraphics*[width=3.4in]{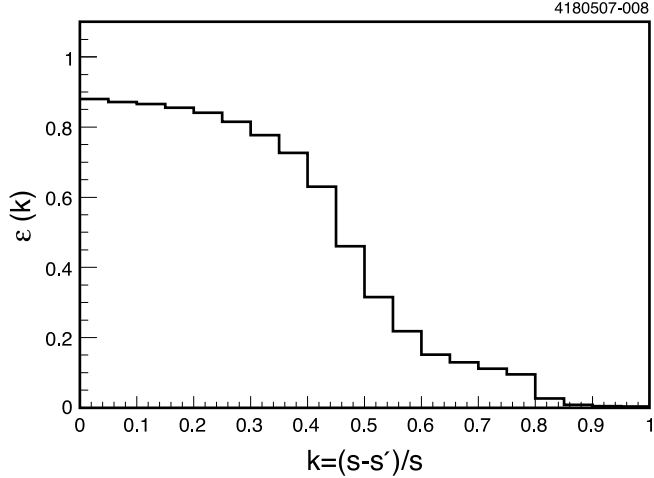}
\caption{Hadronic event detection efficiency as a function of $k$ at 9.996 GeV.
}
\label{FigDiffEff_Ups2sOff}
\end{figure}

Figure~\ref{FigISR_Ups4sOff} shows the $k$-spectra, 
relative decrease in 
squared center-of-mass energy due to initial state radiation, 
The step near $k=0.84$ represents the $c\bar{c}$ threshold.
Table~\ref{tab:effandtrig} summarizes the no-radiation hadronic event 
detection efficiencies $\epsilon(0)$, and Fig.~\ref{FigDiffEff_Ups2sOff} 
shows how the hadronic event detection efficiency varies with $k$.

\section{BACKGROUND}
\label{Background}
\begin{figure}[htbp]
\includegraphics*[width=3.4in]{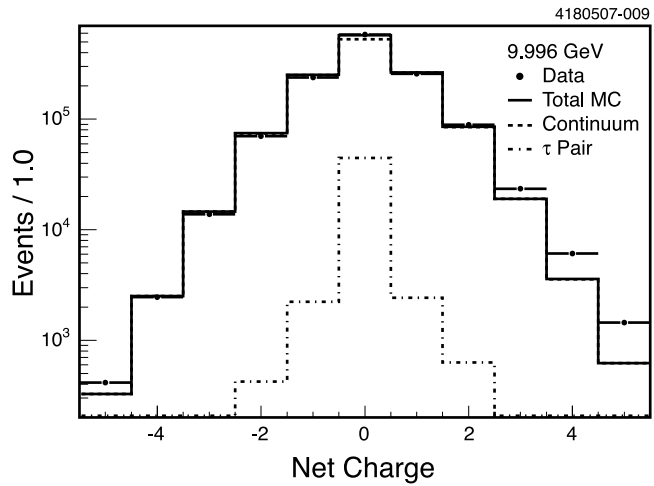}
\caption{ Net charge distribution of events at 9.996~GeV.}
\label{FigNetCharge_Ups2sOff}
\end{figure}

Sources of background were suppressed 
with the help of our hadronic event selection criteria. We deal with  
the remaining level of background events from
two-photon, beam-gas, and 
beam-wall interactions, as well as from $\tau$-pair, $\mu$-pair, 
and Bhabha production in the following.
Figure~\ref{FigNetCharge_Ups2sOff} shows the net charge 
distribution of hadronic events that pass the hadronic event 
selection criteria. 
The Monte Carlo (continuum + $\tau$-pair) describes our data well, however,
there are small discrepancies in the tails (Figs. 1-6). 
These discrepancies may be related 
to the beam-gas or beam-wall interactions not included in our Monte 
Carlo samples. We estimated and subtracted the number of $\tau$-pair events 
using the known QED cross section and the KORALB \cite{KORALB} Monte Carlo simulation. 

The two-photon interactions, which are the  dominant source of 
background after 
 $\tau$-pairs, were studied with a dedicated Monte Carlo that generated
$e^+e^-~\rightarrow~e^+e^-$~+~hadrons events. We found  that the portion of 
two-photon events which survive our selection criteria with a loose cut 
$E_{\mbox{\scriptsize vis}}/2E_{\mbox{\scriptsize beam}} \ge 0.2$
(our nominal cut is 0.5) changes from 1.52\% at 10.538~GeV to 
0.5\% at 6.964~GeV.
For this reason, instead of subtracting the two-photon 
background, we have varied the  $E_{\mbox{\scriptsize vis}}/2E_{\mbox{\scriptsize beam }}$ 
cut to assign an 
additional contribution to the event selection systematic uncertainty in $R$.

\section{RESONANCE EFFECTS AND RADIATIVE CORRECTION}

The beams radiate energy before the annihilation interaction and can then create 
 $c\bar{c}$ and $b\bar{b}$ bound state resonances.
When the resonance decays to hadrons, our observed 
cross section increases. For the purposes of our $R$
measurement, these contributions are sources of background and must be
subtracted. 
The following resonances were considered:  $\Upsilon({\rm3S})$, $\Upsilon({\rm 2S})$, 
$\Upsilon({\rm 1S})$, 
$\psi({\rm 4160})$, $\psi({\rm 3770})$, $\psi({\rm 2S})$, and   $J/\psi$.
To take into account the resonance contributions  we use  Eqs.~(\ref{FormRes})~and~(\ref{sigmaObs}). 
For this purpose we used  the CLEO-measured dielectron widths  of 
$\Upsilon({\rm 1S})$, $\Upsilon({\rm 2S})$, and  $\Upsilon({\rm 3S})$ 
from Ref.~\cite{DiElUpsilon} and 
dielectron widths of $J/\psi$ and $\psi({\rm 2S})$ from Ref.~\cite{DiElPsi}. 
We used dielectron widths of $\psi({\rm 3770})$ and $\psi({\rm 4160})$ 
from the PDG \cite{PDG-04}. 
Table~\ref{tab:res} shows the total cross section of contributing resonances
and the ratio of the resonance contribution to the observed cross section 
for each energy point. We used a sample of low-background charged dipion
transition events in data to check our calculated quarkonium contributions
to our observed hadronic sample. Selecting $\Upsilon({\rm 3S})$,
$\Upsilon({\rm 2S})$, and $\psi({\rm 2S})$ resonance production through
initial state radiation from 10.538~GeV, we studied the following transitions:
$\Upsilon({\rm 3S}) \rightarrow \pi^+\pi^- \Upsilon({\rm 1S})$, 
$\Upsilon({\rm 2S}) \rightarrow \pi^+\pi^- \Upsilon({\rm 1S})$, and 
$\psi({\rm 2S}) \rightarrow \pi^+\pi^- J/\psi $,
where $\Upsilon({\rm 2S})$, $\Upsilon({\rm 1S})$, and $J/\psi$
decay to $\mu^+\mu^-$, to estimate 
resonance contributions at our highest-luminosity point, 
10.538~GeV. We found that our experimental results agreed  
with the theoretical expectation for these modes within 1.5~standard deviations.
\begin{table}[htbp]
\caption{Number of hadronic events after background subtraction, 
radiation and efficiency corrections, total and relative contribution of resonances
through radiative return, and interference effect.}
\label{tab:res}
\begin{center}
\begin{tabular}{cccccc}
\hline\hline
Energy    & Hadronic &$\epsilon(0)\delta_{\mbox{\scriptsize sv}}+I_{\mbox{\scriptsize hard}}$ &  $\sigma_{\mbox{\scriptsize res}}$ & $\sigma_{\mbox{\scriptsize res}}/\sigma_{\mbox{\scriptsize obs}}$ & Interfer- \\
  GeV     & events &       &   nb &  \%  & ence~\% \\    
\hline
   10.538  &  2426337 & 0.935   & 0.056 &  2.08 & -     \\
   10.330  &  ~403743 & 0.935   & 0.044 &  1.62 & 0.14  \\
    9.996  &  1988179 & 0.931   & 0.033 &  1.14 & 0.22  \\
    9.432  &  ~573516 & 0.919   & 0.001 &  0.03 & 0.51  \\
    8.380  &  ~~26705 & 0.901   & 0.003 &  0.09 &   -   \\
    7.380  &  ~~42204 & 0.879   & 0.009 &  0.14 &  -    \\
    6.964  &  ~~14508 & 0.867   & 0.009 &  0.17 &  -    \\ 
\hline\hline
\end{tabular}
\end{center}
\end{table}

Since three runs were at energies just below the Upsilon 
resonance masses, interference between the continuum and the resonance 
can affect the measured continuum cross 
section. The magnitude and sign of this effect depends on 
the difference between the center-of-mass energy and the resonance mass.
The CESR beam energy spread, 
which is of order  4~MeV,
convolves the interference effects from different center-of-mass energies, 
yielding an overall negative effect on the measured cross section below resonance.
In the last column of Table~\ref{tab:res} are
listed calculated corrections to the  continuum cross section 
to compensate  upward for the effect of interference and beam 
energy spread \cite{DiElUpsilon}.  

We applied the
procedure described in Section~\ref{ANALYSIS-OVERVIEW} by
evaluating the 
$\epsilon(0)\delta_{\mbox{\scriptsize sv}}+I_{\rm hard}$ term of Eq.~(\ref{FormCross}) 
to calculate the radiative correction for each 
incident energy point. The efficiencies at different values of $k$ 
(see Fig.~\ref{FigDiffEff_Ups2sOff}) from  continuum Monte Carlo have 
been used to calculate $I_{\mbox{\scriptsize hard}}$. 
Table~\ref{tab:res} shows the radiation and 
efficiency factor [see Eq.~(\ref{FormCross})] corrections for each energy. 
The decrease of $\epsilon(0)\delta_{\mbox{\scriptsize sv}}+I_{\rm hard}$ 
with energy mainly comes from the efficiency decrease 
(see Table~\ref{tab:effandtrig}).

\section{SYSTEMATIC UNCERTAINTIES AND RESULTS}
The following have been considered as sources of systematic 
uncertainty for each continuum cross 
section measurement: luminosity, radiative correction, 
trigger efficiency for hadronic events,
multiplicity correction, and hadronic event selection criteria
(see Table \ref{tab:systematic}). 

The hadronic vacuum polarization terms in Eq.~(\ref{deltaVP}) are
of the order of 0.01. Since they dominate the uncertainty in the radiative 
correction, we have assigned a $1\%$ uncertainty in $\sigma_0$ from 
this source.

Each selection criterion, except for multiplicity, has independently
been varied by about 20\% to estimate its systematic uncertainty.
Moreover, we used a visible energy  $E_{\rm vis}/2E_{\rm beam}$ 
cut variation of about 40\% to assign an additional systematic 
uncertainty related to the two-photon background and the visible 
energy shift: we assign half the difference between $R$ values 
obtained with cuts on $E_{\rm vis}/2E_{\rm beam}$ of 0.3 and 0.4 
(0.6 and 0.7) as systematic uncertainty associated
with two-photon background (visible energy shift).

Even though the multiplicity cut was 
part of  the hadronic event selection criteria, we 
assigned a separate systematic uncertainty 
to our correction procedure by taking 
the difference between the corrected and uncorrected $R$ 
values at the nominal cut level. 
The summary of all systematic uncertainties is given in 
Table~\ref{tab:systematic}. At each energy  we divide the systematic uncertainty
into a common uncertainty that correlated across all energy points and an
uncorrelated uncertainty  that is independent for each energy point.
The decrease of the uncorrelated systematic uncertainty with 
decreasing beam energy is mainly due to the energy dependence of the 
two-photon interaction background cross section.  
\begin{table}[htbp]
\caption{Systematic uncertainty (in \%) assigned to each energy point. 
The last two rows show common and uncorrelated uncertainties.}  
\label{tab:systematic}
\begin{center}
\begin{tabular}{lccccccc}
\hline\hline
 Energy~(GeV)       & 10.538  & 10.330 & 9.996 & 9.432 & 8.380 & 7.380 & 6.964  \\
\hline 
   Luminosity       & 1.00 & 1.10 &  1.10 & 1.10 & 0.90 & 0.90 & 1.00\\
  Trigger           & 0.09 & 0.09 &  0.11 & 0.08 & 0.12 & 0.13 & 0.19\\
  Radiative         & 1.00 & 1.00 &  1.00 & 1.00 & 1.00 & 1.00 & 1.00\\
 correction         &      &      &       &      &      &      &     \\
  Multiplicity      & 1.06 & 1.38 &  0.99 & 0.84 & 0.43 & 0.38 & 0.38\\
correction          &      &      &       &      &      &      &     \\
  Hadron event      & 1.51 & 1.09 &  1.31 & 1.31 & 1.05 & 1.02 & 0.79\\
selection           &      &      &       &      &      &      &     \\ 
\hline
Total               & 2.32 & 2.30 &  2.21 & 2.15 & 1.76 & 1.74 & 1.68\\
\hline
Common              & 1.87 & 1.67 &  1.85 & 1.87 & 1.62 & 1.64 & 1.58\\
Uncorrelated        & 1.37 & 1.59 &  1.22 & 1.05 & 0.70 & 0.57 & 0.55\\
\hline
\hline
\end{tabular}
\end{center}
\end{table}
\begin{figure}[htbp]
\begin{center}
\includegraphics*[width=3.4in]{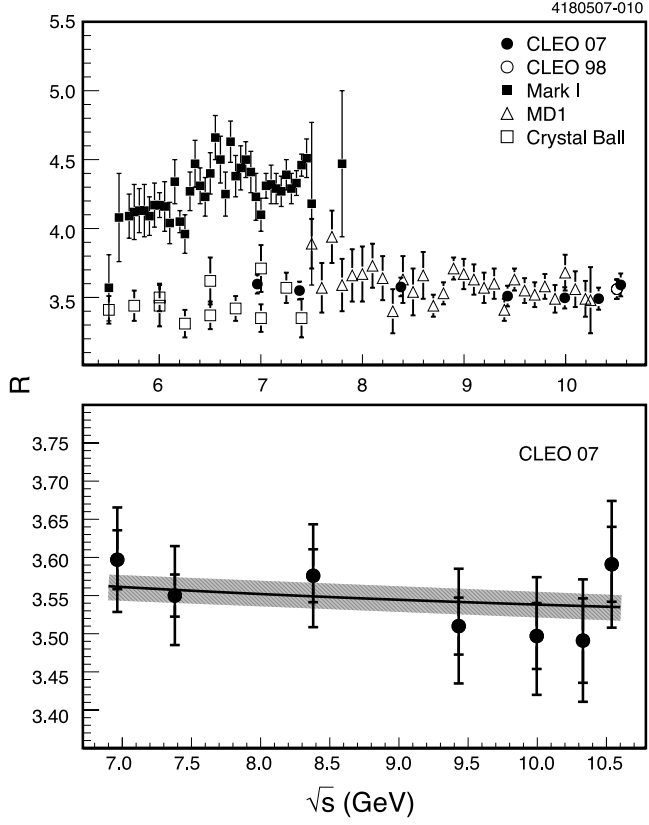}
\caption{R measurement at 5.0 - 10.5 GeV c.m. energies. The top figure compares our measurement 
with other experiments \cite{PDG-04}. The bottom plot shows our new 
results,where two sets of uncertainties 
represent combined uncorrelated and statistical uncertainties and total uncertainties; the line represents Eq.~(\ref{FormR}) 
with our average $\Lambda=0.31$~GeV, and the 
shaded area
indicates the $R$-values corresponding to one standard deviation in the
uncorrelated systematic uncertainty in $\Lambda$.}
\label{FigRvalues}
\end{center}
\end{figure}

Values of $R$  for each of the energy points are shown in 
Table~\ref{tab:R} 
and are presented in  Fig.~\ref{FigRvalues} along with MARK~I~\cite{MARKI}, 
MD1~\cite{MD1}, Crystal Ball~\cite{CrystalBall}, 
and prior CLEO~\cite{CLEO:R} results. 
\begin{table}[htbp]
\caption{Measured values of $R(s)$  with statistical and systematic 
(common and uncorrelated) uncertainties, respectively.} 
\label{tab:R}
\begin{center}
\begin{tabular}{|c|c|}
\hline\hline
 $\sqrt s$ &  $R(s)$  \\ 
    GeV    &          \\ 
\hline
   ~~~10.538~~~      & ~~~$3.591~\pm~0.003~\pm~0.067~\pm~0.049$~~~  \\ 
      10.330      & $3.491~\pm~0.006~\pm~0.058~\pm~0.055$  \\ 
      ~9.996      & $3.497~\pm~0.004~\pm~0.064~\pm~0.043$  \\ 
      ~9.432      & $3.510~\pm~0.005~\pm~0.066~\pm~0.037$  \\ 
      ~8.380      & $3.576~\pm~0.024~\pm~0.058~\pm~0.025$  \\ 
      ~7.380      & $3.550~\pm~0.019~\pm~0.058~\pm~0.020$  \\ 
      ~6.964      & $3.597~\pm~0.033~\pm~0.057~\pm~0.020$  \\ 
\hline\hline
\end{tabular}
\end{center}
\end{table}
Our new measurements of $R$  are in good agreement with previous MD1, 
Crystal Ball, and CLEO measurements; however, they do not agree with 
the MARK~I results.

Table~\ref{tab:alpha} shows the $\alpha_s$ values obtained by solving Eq.~(\ref{FormR}) 
(with four flavors) 
for $\alpha_s$ at each of the seven energies 
(extraction of $\alpha_s$ when quark masses are taken into account can be found in reference \cite{Kuhn}). 
Comparing  $\alpha_s$ values with the QCD
predictions \cite{Bethke-07} at our energies, which assumes the combined world average 
of $\alpha_s(M_{Z}^2) = 0.1189\pm0.0010$, 
 we find agreement within our quoted uncertainties. 
Moreover, we compare our results at 10.330, 9.996, and 
9.432~GeV energies with CLEO measurements \cite{CLEO-direct-photon} at $\Upsilon({\rm1S})$, 
$\Upsilon({\rm 2S})$, and $\Upsilon({\rm 3S})$ energies, which are based on direct photon 
spectrum studies, and  we find  a good agreement within our quoted uncertainties as well.
\begin{table}[hbtp]
\caption{Derived values of $\alpha_s(s)$ with statistical and systematic 
(common and uncorrelated) uncertainties, respectively using the prescription outlined in the text.} 
\label{tab:alpha}
\begin{center}
\begin{tabular}{|c|c|}
\hline\hline
 $\sqrt s$ &  $\alpha_{s}(s)$\\
    GeV    &                 \\    
\hline
~~~10.538 ~~~  &   ~~~$0.232~\pm~0.003~\pm~0.061~\pm~0.045$~~~ \\
   10.330      &  $0.142~\pm~0.005~\pm~0.051~\pm~0.049$ \\
    ~9.996     &  $0.147~\pm~0.004~\pm~0.057~\pm~0.038$ \\
    ~9.432     &  $0.159~\pm~0.004~\pm~0.058~\pm~0.033$ \\
    ~8.380     &  $0.218~\pm~0.022~\pm~0.053~\pm~0.023$ \\
    ~7.380     &  $0.195~\pm~0.017~\pm~0.052~\pm~0.018$ \\
    ~6.964     &  $0.237~\pm~0.030~\pm~0.052~\pm~0.018$ \\ 
\hline\hline
\end{tabular}
\end{center}
\end{table}

To test the compatibility with other measurements of $\alpha_s$ we note the
expected running of $\alpha_s$ with energy \cite{PDG-04}:
\begin{eqnarray}
\alpha_s(s) & = & \frac{4\pi}{\beta_0\ln(s/\Lambda^2)}\left[ 1 - \frac{2\beta_1}{\beta_0^2}\frac{\ln[\ln(s/\Lambda^2)]}{\ln(s/\Lambda^2)} 
\right. \nonumber \\
            &   & \left. \mbox{}+\frac{4\beta_1^2}{\beta_0^4\ln^2(s/\Lambda^2)}\times\left(\left(\ln\left[\ln(s/\Lambda^2)\right]-\frac{1}{2}\right)^2 
\right. \right. \nonumber \\
            &   & \left.\left. \mbox{}+\frac{\beta_2\beta_0}{8\beta_1^2}-\frac{5}{4}\right)\right] ,
\label{AlphaS}
\end{eqnarray}
where $n_f$ presents the number of quarks which have mass less than $\sqrt s$/2, 
$\Lambda$ represents the QCD energy scale, and the $\beta$-functions are 
defined as follows: 
$\beta_0=11-2n_f/3$, $\beta_1=51-19n_f/3$, and $\beta_2=2857-5033n_f/9+325n_f^2/27$.

To find $\Lambda$, we use our $\alpha_s$ values at each energy point and
solve Eq.~(\ref{AlphaS}), assuming $n_f$ is equal to 4. 
The value of $\Lambda$  varies from  0.11 at 10.330~GeV to 0.67
at 10.538~GeV. Using Eq.~(\ref{AlphaS}) with our average 
value of  $\Lambda$, we extract the value of  the $\alpha_s$ at ${\sqrt s}=M_Z$.
Our results for $\alpha_s$ imply 
$\Lambda = 0.31^{+0.09+0.29}_{-0.08-0.21}~\rm{GeV}$
and 
$\alpha_s(M_{Z}^2) = 0.126\pm0.005~^{+0.015}_{-0.011}$, 
where the uncertainties represent
statistical and total systematic, respectively.  

Our results for $\alpha_s(M_{Z}^2)$ and $\Lambda(n_f=4)$ agree with the world averages 
$\alpha_s(M_{Z}^2)=0.1189 \pm 0.0010$ \cite{Bethke-07} and 
$\Lambda(n_f=4)=0.29 \pm 0.04$~GeV \cite{Bethke-04}.

In conclusion, we have shown that our measurements of $R(s)$  are consistent
with the QCD predictions of Eq.~(\ref{defR}) and Eq.~(\ref{AlphaS}) with better 
precision than previous measurements of $R(s)$.  However, since the Eq.~(\ref{defR}) dependence of
 $R(s)$ on $\alpha_s(s)$ is only in higher  order terms, we do not have a
 determination of $\alpha_s(s)$  or $\Lambda$ that is competitive with measurements
 by other methods.

We gratefully acknowledge the effort of the CESR staff
in providing us with excellent luminosity and running conditions.
D.~Cronin-Hennessy and A.~Ryd thank the A.P.~Sloan Foundation.
This work was supported by the National Science Foundation,
the U.S. Department of Energy, and
the Natural Sciences and Engineering Research Council of Canada.


\begin{thebibliography}{99}
\bibitem{first20}
{Bonneau and Martin \cite{BonneauMartin} divide the radiated energy spectrum 
between  $\sigma_{\mbox{\scriptsize sv}}$ and $\sigma_{\mbox{\scriptsize hard}}$ at  $k_{0}\ll1$. 
For convenience, Behrends and Kleiss \cite{BehrendsKleiss} (and we)    remove $1-t\ln k_{0}$ 
from $\sigma_{0}$ and add it to $\sigma_{\rm hard}$ by extending the hard integral down 
to $k=0$.}
\end{thebibliography}
\end{document}